\documentclass{article}

\usepackage[preprint, nonatbib]{neurips_2020}

\usepackage[utf8]{inputenc} 
\usepackage[T1]{fontenc}    
\usepackage{hyperref}       
\usepackage{url}            
\usepackage{booktabs}       
\usepackage{amsfonts}       
\usepackage{nicefrac}       
\usepackage{microtype}      

\usepackage{graphicx}
\usepackage{amsmath}
\usepackage{float}
\usepackage{longtable}
\usepackage{lscape}
\usepackage{bigstrut}

\title{Incorporating network based protein complex discovery into automated model construction}

\author{%
   Paul Scherer   \quad\quad
   Maja Tr\c{e}bacz    \quad\quad
   Nikola Simidjievski  \\
   \textbf{Zohreh Shams \quad \quad
   Helena Andres Terre\quad\quad
   Pietro Li\`o\quad\quad
   Mateja Jamnik} \\ 
   Department of Computer Science and Technology,\\ University of Cambridge, UK 
}

\begin{document}

\maketitle

\begin{abstract}
We propose a method for gene expression based analysis of cancer phenotypes incorporating network biology knowledge through unsupervised construction of computational graphs. The structural construction of the computational graphs is driven by the use of topological clustering algorithms on protein-protein networks which incorporate inductive biases stemming from network biology research in protein complex discovery. This structurally constrains the hypothesis space over the possible computational graph factorisation whose parameters can then be learned through supervised or unsupervised task settings. The sparse construction of the computational graph enables the differential protein complex activity analysis whilst also interpreting the individual contributions of genes/proteins involved in each individual protein complex. In our experiments analysing a variety of cancer phenotypes, we show that the proposed methods outperform SVM, Fully-Connected MLP, and Randomly-Connected MLPs in all tasks. Our work introduces a scalable method for incorporating large interaction networks as prior knowledge to drive the construction of powerful computational models amenable to introspective study.
\end{abstract}

\section{Introduction}
Gene expression data is commonly used within research intersecting cancer data research and machine learning as it is seen as a crucial component towards understanding the molecular status of tumour tissue. In its most common form an observation of gene expression data is presented as an $n$-dimensional features vector of continuous values where each element of the vector corresponds to the expression level of a particular gene in the sample. Classically, this representation is directly used to learn a prediction model for tasks such as cancer disease subtype classification or as part of a larger system integrating data from multiple modalities \cite{canceraivae, Esteva2019}. 

The high dimensionality and noisiness of the gene expression data poses significant problems to learning algorithms. This causes models to overfit, learn noise, and fail to capture any biologically relevant information. As a result, practitioners commonly aim to constrain model complexity by incorporating various approaches for regularisation including dimensionality reduction and use of prior biological knowledge to inductively bias models towards learning representations with favourable characteristics \cite{Dutil2018TowardsGE, canceraivae, Simon13asparse-group, networkregularization, Cawley, Min2018}. A part of this research on using prior knowledge focuses on the incorporation of gene interaction networks as external priors into the predictive model to guide the learning process. The overall goal of applying network-based analysis to personal genomic profiles is to identify network modules that are both informative of cancer mechanisms and predictive of cancer phenotypes. A survey of such approaches is covered in Zhang et al. \cite{Zhang2017}.

In this work we utilise topological clustering algorithms chiefly used for the identification of protein complexes and functional modules within PPI networks to define the structure of computational graphs in an unsupervised manner. This deterministic procedure produces sparse computation graphs which relates genes to named protein complexes, structurally parameterising individual functions for the "activity" of each complex based on an input gene expression profile. Further connecting the complex activities to cancer phenotypes defines a supervised predictive model which analyses the activity patterns of higher level functional modules (protein complexes) to cancer phenotypes. Our approach effectively constrains the hypothesis space of models via structural biases obtained through unsupervised analyses of network biology entities. Figure \ref{fig:geo2dr_example} in Appendix A features a simplified diagram of this process over an input genomic profile dataset and a toy interaction network used to construct the topology of the computational graph.

\section{Methods}
The proposed method, which we will call \textit{PComplexNet}, incorporates prior biological knowledge imbued within the structure of supplied PPI networks and protein complexes discovered via topological clustering algorithms to construct a bipartite graph between genes/proteins and functional modules. This bipartite graph serves as the structural foundation of the computational graphs that will be further augmented into predictive models for cancer phenotype. Crucially, this means that the structure of the output computational graphs is defined in a purely unsupervised and deterministic manner over external curated knowledge. 

The procedure for constructing the computational graphs is best described in three stages: (i) obtaining a study specific subgraph of the PPI network, (ii) discovering protein complexes that serve as higher level features, and (iii) constructing the factor and computational graphs. 

\subsection{Processing input data and external PPI network to generate study PPI network.}
Let us assume an input gene expression dataset $\mathbf{X} \in \mathbb{R}^{m \times k}$ describing $m$ patient observations with $k$-dimensional vectors of gene expression values. Furthermore let us assume an external PPI network $\mathcal{G}_{\mathrm{PPI}} = (V_{\mathrm{PPI}}, E_\mathrm{PPI})$, such as one from the STRING-DB 9606 Homo Sapiens PPI network \cite{stringdb}. For our purpose, this PPI network is an unweighted graph with nodes ($ V_{\mathrm{PPI}}$) labeled by the names of proteins, and no additional node or edge features. We induce a subgraph of the input network $\mathcal{G}_{\mathrm{S}} \subseteq \mathcal{G}_{\mathrm{PPI}} $. The nodes of $\mathcal{G}_{\mathrm{S}}$ are the intersection of the common $k$ genes in the input gene expression dataset $\mathbf{X}_{\mathrm{genes}}$ and genes in the PPI network; in other words $V_{\mathrm{S}} = \mathbf{X}_{\mathrm{genes}} \cap V_{\mathrm{PPI}}$. The induced subgraph $\mathcal{G}_{\mathrm{S}} = (V_{\mathrm{S}}, E_{\mathrm{S}})$ is the graph whose vertex set is $V_{\mathrm{S}}$ and whose edge set consists of all of the edges in $E_\mathrm{PPI}$ that have both endpoints in $V_{\mathrm{S}}$. This action is illustrated in the top row of actions in Figure 1. We denote $\mathcal{G}_{S}$ our \textit{study PPI network} since it is the "cut out" of the external PPI network relevant to our study.

\subsection{Protein complex discovery}
Given the induced study subnetwork, we use a topological clustering algorithm $\mathcal{C}$ such as DPCLUS \cite{dpclus} to discover protein complexes within the study PPI network $\mathcal{G}_{\mathrm{S}}$. The aim of the clustering algorithms is to discover protein complexes represented as a set of induced subgraphs $\mathcal{C}(\mathcal{G}_{\mathrm{S}}) = \{c_1, c_2, \ldots, c_l\}$, where $l$ is the number of complexes discovered by $\mathcal{C}$. The number of protein complexes found, $l$, is not dependent on the user, but rather the application of the clustering algorithm $\mathcal{C}$ upon the input study network.

It is worth noting that we specifically chose clustering algorithms that do not partition the graph. In other words, a single protein may be part of multiple complexes. This is to reflect the fact that proteins may be involved in several biological processes and complexes. Another note to make is that not all proteins in $\mathcal{G}_{\mathrm{S}}$ will necessarily be assigned to clusters by $\mathcal{C}$. We are not arbitrarily forcing all genes to be part of our constructed models, and this acts as a form of feature selection upon the input $\mathbf{X}$ by $\mathcal{C}(\mathcal{G}_{\mathrm{S}})$.

\begin{figure}
    \includegraphics[width=\textwidth]{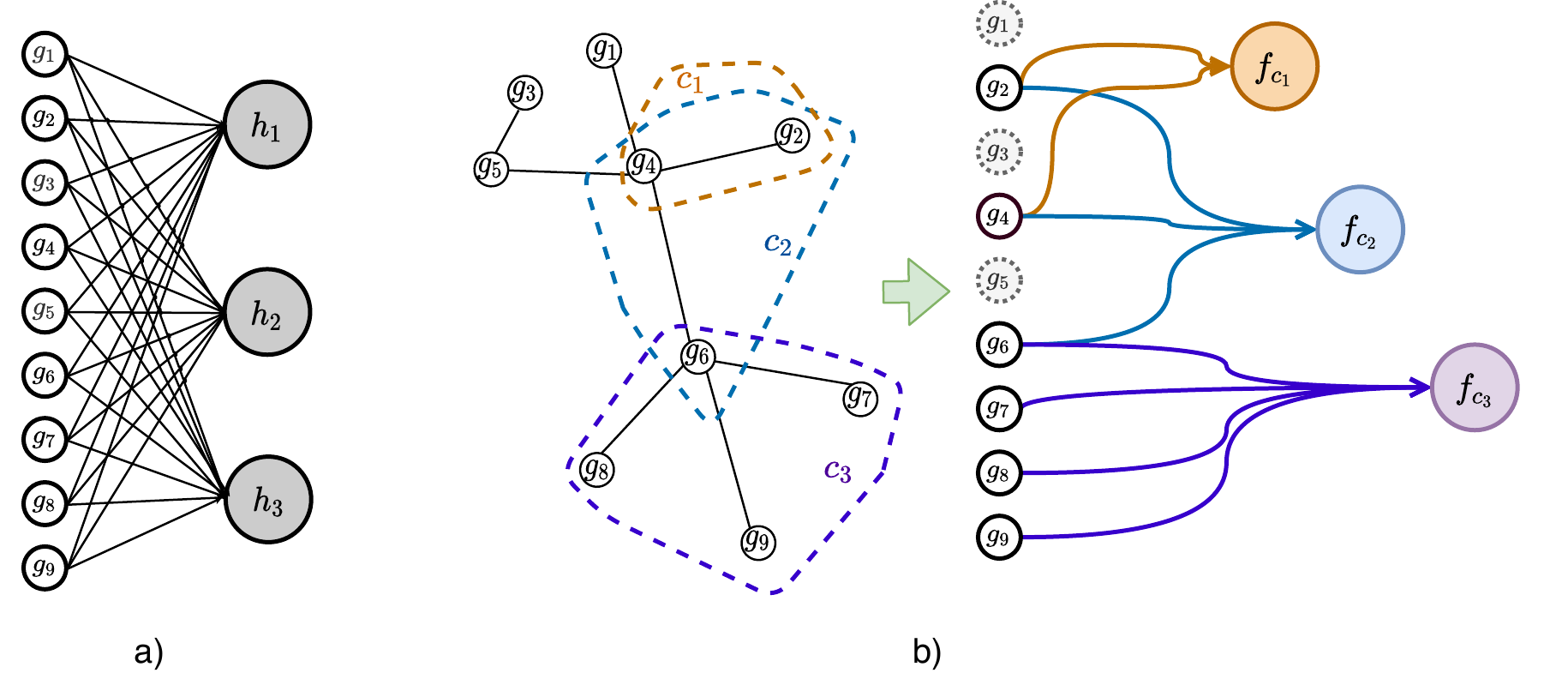}
    
    \caption{This figure depicts a side-by-side comparison of a) a typical Fully-Connected MLP and b) the factor graph produced through PComplexNet. The factor graphs produced through PComplexNet are considerably sparser and incorporate biological knowledge from the PPI network and protein complexes discovered within. The input features used in the model are cut down through two steps. The first set of genes removed from the extraction of the study network $\mathcal(G)_{S}$. The second set of features are removed through the clustering process $\mathcal{C}(\mathcal{G}_{S})$.}
    \label{fig:2}
\end{figure}


\subsection{Computational graph construction and predictive models}

The output of the clustering algorithm $\mathcal{C}(\mathcal{G}_{\mathrm{S}}) = \{c_1, c_2, \ldots, c_l\}$ enables the construction of a bipartite factor graph. Herein, each of the protein complexes is assigned a uniquely labelled node $c_i$ and each protein within the set of proteins involved in one or more complexes is also given a labeled node by their name. Directed edges link proteins to complexes $c_i$ they are a member of. This construction gives the factorisation of a parametric function $f_{c_i}: c_i \rightarrow \mathbb{R}$ computed from the proteins involved in $c_i$. The function $f_{c_i}(\cdot)$ can be set by the practitioner or learned through a neural network. 

The parameterisations $f_{c_i}: c_i \rightarrow \mathbb{R}$ in our proposal is a stark contrast to arbitrarily chosen hidden-state activations $h_{i}: \mathbb{R}^{k} \rightarrow \mathbb{R}$ found in conventional application of fully-connected multi-layer perceptrons. Firstly, each of the $c_i$ denotes a protein complex activity, a biologically relevant structure modelled through incorporation of external PPI and topological clustering algorithm, instead of an arbitrarily chosen hidden state node. The proteins that are members of $c_i$, and only those proteins, affect its activity level $f_{c_i}: c_i \rightarrow \mathbb{R}$, instead of all input features. This is a strong and explicit inductive bias if $f_{c_i}$ is learned through a neural network. 


We construct computational graphs for cancer phenotype prediction by further augmenting the current gene/protein to protein complex factor graph to include complete connections between the protein complexes $c_i$ to target nodes gained when encoding the target observations $\mathbf{Y}$. As such each $f_{c_i}: c_i \rightarrow \mathbb{R}$ computing the individual protein complex "activity" is learned over minimising the global cross-entropy loss between gene expression values and the target phenotypes.

\section{Empirical evaluation}

In order to evaluate the proposed method for model construction, we used publicly available gene expression data from the METABRIC Breast Cancer Consortium (METABRIC) \cite{curtis2012genomic} and The Cancer Genome Atlas Head-Neck Squamous Cell Carcinoma (TCGA-HNSC) \cite{Rendleman2019, tcgahnsc}. Using the former we evaluate on three classification tasks of predicting:  Distance Relapse (binary classification), PAM50 breast tumour cancer subtypes (5-class classification), Integrative Cluster (IC10) subtypes (11-class classification). Using the latter we evaluate on two classification tasks of tumour grade (4-class classification) and 2 year relapse free survival (binary classification). All classification tasks were evaluated by mean percentage accuracy over a stratified 5-fold cross-validation. The specific details about the datasets, experimental setup, and methods are given in Appendix B. 

Amongst the considered methods are: support vector machine with RBF kernel (SVM), a Fully-Connected (FC) two layer neural network with 1600 hidden layer nodes\footnote{This number of hidden nodes was chosen to closely match the number of protein complexes used in PComplexNet + DPCLUS, the best performing of the proposed methods.}, Randomly-Connected (RC) MLP, and our proposed model constructor coupled with a variety of topological clustering algorithms. Each of our models is referred to as PComplexNet + $\mathcal{C}$, where $\mathcal{C}$ refers to one of: MCODE \cite{mcode}, COACH \cite{coach}, IPCA \cite{ipca}, or DPCLUS \cite{dpclus} clustering algorithms. The hyperparameters of the clustering algorithms were set to their default values.

The main comparative results are summarised in Table~\ref{tab:addlabel} for the METABRIC and TCGA-HNCS datasets. The results show that all variations of the computational graphs produced by PComplexNet (regardless of the clustering algorithms) outperform both the SVM and Fully-Connected MLP baselines. More specifically, PComplexNet + DPCLUS considerably outperform the baselines on all five classification tasks, making especially substantial gain in IC10 subtype prediction in the case of METABRIC. We attribute these performance gains of PComplexNet over Fully-Connected MLPs to two related advantages. Firstly, PComplexNet's sparser model complexity allows more "weight" to be assigned to each of the input signals used. Similarly, the sparse connectivity also helps generalisability in a similar way to the dropout regularisation method. However, in contrast the connectivity is set, explicit, and realised through incorporation of prior knowledge rather than being random and ephemeral. This brings us to the second advantage of PComplexNets --- the structure of the computational graphs, and thus the representations, explicitly incorporate biological knowledge of protein complex membership as intermediate states. In other words, they are not "hidden" nodes. The learned activities of the protein complexes are explicitly factorised to the gene expression measurements of the genes/proteins that have a membership in the complex. 

To show that PComplexNet benefits from both of these advantages, and not only from the first advantage of regularisation via sparse connections, we show that the performance of PComplexNet + DPCLUS also outperform computational graphs constructed through a random process (RC MLP). The differing performances on the choice of clustering algorithm $\mathcal{C}(\cdot)$ reflects the different assumptions made by researchers on what topological structures within $\mathcal{G}_{S}$ contain protein complexes. MCODE and DPCLUS exhibit stricter rules on complex candidates with fewer, smaller, and more tightly knit clusters than either COACH or IPCA. This may be interpreted as these two methods constraining the hypothesis space more and incorporating "more" expert knowledge which is helpful to the classification tasks. Naturally PComplexNet is agnostic to the choice of $\mathcal{C}(\cdot)$, therefore various combinations or set complexes may be explored in further work.

\begin{table}[t]
  \centering
 \caption{Average accuracy of stratified 5-fold cross-validation using all of the gene expression features of METABRIC and TCGA HNSC. }
    \resizebox{\textwidth}{!}{
    \begin{tabular}{l|rrr|rr}
    \hline
          & \multicolumn{3}{c|}{METABRIC} & \multicolumn{2}{c}{TCGA-HNSC} \bigstrut[t]\\
          & \multicolumn{1}{c}{DR} & \multicolumn{1}{c}{PAM50} & \multicolumn{1}{c|}{IC10} & \multicolumn{1}{c}{Tumour Grade} & \multicolumn{1}{c}{2 Year RFS} \bigstrut[b]\\
    \hline
    SVM   & 59.39 + 9.23 & 75.84 + 2.07 & 68.28 + 2.83 & 56.61 + 3.67 & 56.92 + 5.32 \bigstrut[t]\\
    FC MLP & 64.24 + 3.86 & 76.03 + 2.62 & 66.11 + 3.12 & 59.03 + 1.49 & 55.96 + 5.52 \\
    RC MLP &  65.30 + 1.04 &  75.87 + 1.60 & 67.76 + 3.14 &  54.67 + 2.88 & 57.07 + 1.33\\
    PComplexNet + MCODE & 66.06 + 1.92 & 77.10 + 1.61 & 67.92 + 3.56 & 56.42 + 2.49 & 58.65 + 7.91 \\
    PComplexNet + COACH & 57.27 + 3.14 & 76.89 + 2.89 & 71.56 + 2.01 & 57.84 + 2.49 & 56.15 + 2.48 \\
    PComplexNet + IPCA & 65.05 + 3.42 & 77.71 + 2.27 & 68.88 + 5.05 & 55.82 + 2.34 & 55.58 + 2.94 \\
    PComplexNet + DPCLUS & \textbf{67.42 + 2.25} & \textbf{79.53 + 2.46} & \textbf{76.62 + 1.59} & \textbf{62.64 + 2.12} & \textbf{59.04 + 5.21} \bigstrut[b]\\
    \hline
    \end{tabular}%
    }
  \label{tab:addlabel}%
\end{table}%

\section{Conclusion}

We presented PComplexNet, a scalable unsupervised approach to incorporating biological knowledge embedded in the structure of PPI networks for automated construction of computational graphs for genome analysis. PComplexNet has several distinguishing properties. First, it provides a biologically relevant mechanism for model regularisation, resulting in structurally constrained models that yield better predictive performance. Second, PComplexNet is scalable and readily applicable to other genomic data analysis tasks. For example, the computational graphs can be seamlessly incorporated into larger integrative frameworks handling multiple modalities such as the integrative variational auto-encoders \cite{canceraivae}. Finally, there is no arbitrary decision making on the number of hidden nodes or their biological relevance as in standard MLPs. Each node within our computational graphs is either a gene, a phenotype, or a protein complex. The structure describes a knowledge-directed factorisation of the parametric function for the activity of a protein complex based on the expression levels of its constituent gene/proteins. This makes introspective study into the individual contributions of entities in the model and patterns as a whole more amenable.





\newpage

\bibliographystyle{unsrt} 
\bibliography{main} 

\newpage
\appendix

\section{Diagram of PComplexNet constructions}
\begin{figure}[H]
    \centering
    \includegraphics[width=\textwidth]{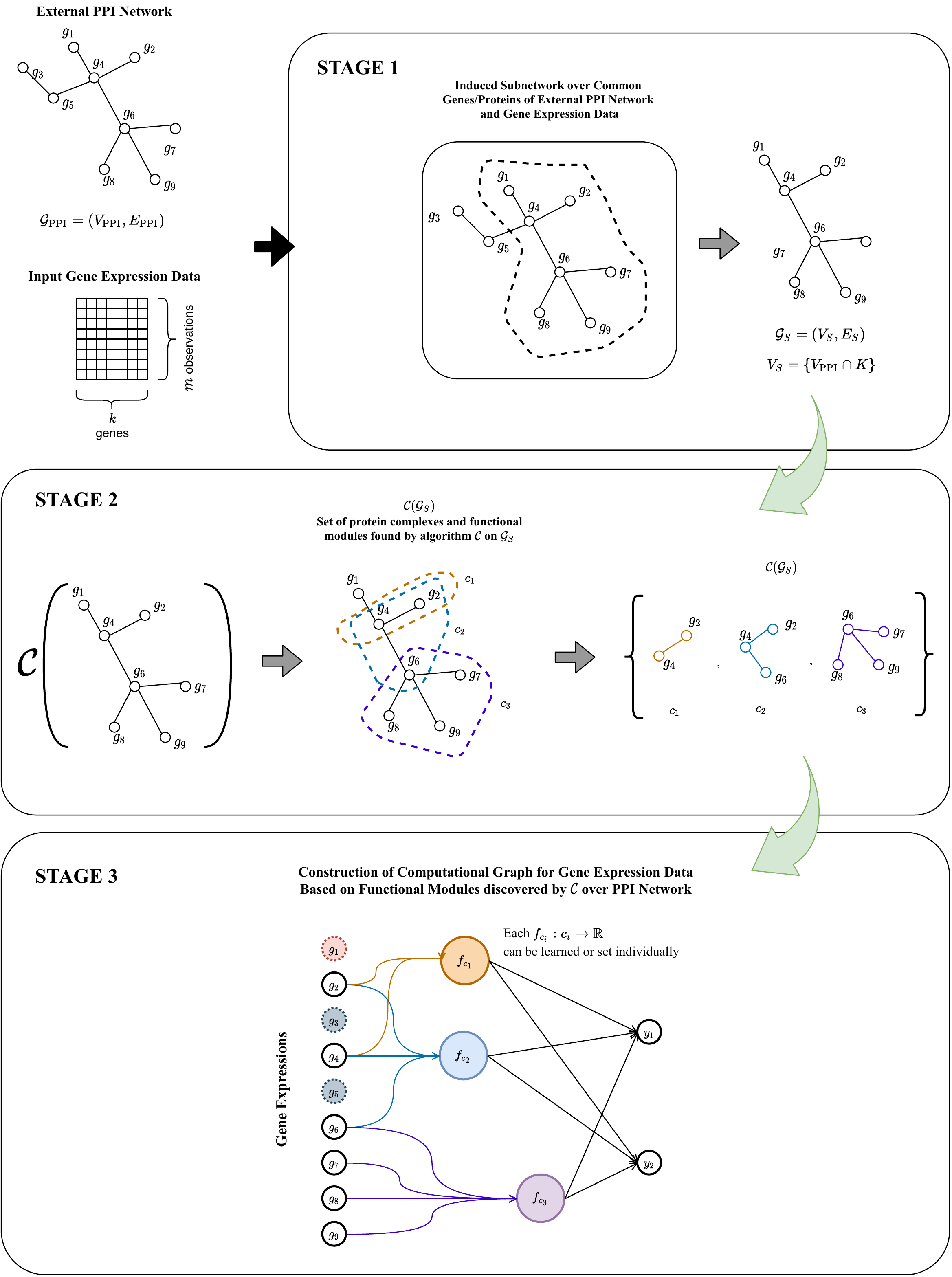}
    \caption{An overview of our procedure for incorporating PPI network based protein complex discovery and constructing computational graphs for gene expression analysis. Each row corresponds to a distinct stage of the procedure detailed in Section 2.}
    \label{fig:geo2dr_example}
\end{figure}

\newpage
\section{Data and experimental setup}
Our first dataset consists of the mRNA expression data and clinical data of breast cancer patient samples in the METABRIC cohort \cite{curtis2012genomic}. Herein we tackle several classification tasks over the 1980 breast cancer patients, representing a particularly large dataset. Each observation is represented by 24368 dimensional vector corresponding to the expression values of measured genes. We evaluate the predictive performance over the proposed methods ability to correctly predict:

\begin{itemize}
    \item Distance relapse (DR), a binary classification task. This target describes whether or not the cancer has metastasised to another organ after initial treatment.
    \item IntegrativeCluster subtypes \cite{curtis2012genomic} (IC), a 11 class prediction task.
    \item PAM50 breast tumour cancer subtype \cite{Prat2010} (PAM50), a 5 class prediction task for luminal A, luminal B, HER2-Enriched, basal-like and claudin-low tumour subtypes.
\end{itemize}

The second dataset under consideration is The Cancer Gene Atlas Head and Neck Squamous Cell Carcinoma dataset (TCGA-HNSC). TCGA-HNSC mRNA expression data analysed in this study was obtained through the National Cancer Institute Genomic Data Commons Data Portal, https://portal.gdc.cancer.gov/ as in Rendleman et al. \cite{Rendleman2019}. The dataset contains 528 TCGA-HNSC cases wherein we focus on the 20501 RNA expression variables (dropping 30 genes with missing values in the dataset). The clinical targets include:

\begin{itemize}
    \item Tumour grade, a 4 class prediction task.
    \item 2 Year Relapse Free Survival (2 Year RFS), a binary prediction task.
\end{itemize}

The class distribution of each of the targets is heavily skewed within the datasets. Hence, we evaluate the methods on a 5-fold cross-validation with class stratified train-test splits and record the mean percentage accuracy.

Amongst the considered methods are: support vector machine with RBF kernel (SVM), a Fully-Connected (FC) two layer neural network with 1600 hidden layer nodes, Randomly-Connected (RC) MLP, and our proposed model constructor coupled with a variety of topological clustering algorithms. Each of our models is referred to as PComplexNet + $\mathcal{C}$, where $\mathcal{C}$ refers to one of: MCODE, COACH, IPCA, and DPCLUS clustering algorithms. The hyperparameters of the clustering algorithms were set to their default values.

The Fully-Connected MLP and the computational graphs of PComplexNet were trained through optimisation of the cross entropy loss. The loss was optimised using Adam with a mini batch size of 32 and 250 epochs and an initial learning rate of 0.001. The weight parameters were initialised using the Xavier uniform initialisation.

As the structure of the computational graphs is driven largely by the structure of the external PPI network and the number/members of the protein complexes discovered, we require a sanity check to see that it actually captures any biologically relevant information. Naturally, the structure of the PPI network itself is explained and justified by the maintainers/proposers/curators of the databases. Similarly, the biological relevance of the clustering algorithms used upon the PPI networks is also reasoned and justified within each of the original papers. Therefore, it is safe to assume these components of PComplexNet as trustworthy. Hence, our task here is to find whether the computational graphs constructed through PComplexNet obtain better scores than the SVMs and FC-MLP because the structure and learned activity functions capture meaningful biological relationships. 

In order to test this we construct computational graphs with random number of "discovered protein complexes" and random number of connections describing the protein memberships to clusters. The random numbers are drawn from a uniform distribution between $l \in [30, 12000]$ for the number of protein complexes\footnote{This range was chosen to roughly reflect the number of protein complexes found in the chosen clustering algorithms on the STRING-DB PPI network} and $u \in [1, l*k]$ random protein to complex connections. For an empirical evaluation, 100 instances of such random computational graphs were constructed to obtain a Monte Carlo aggregate mean score shown in Table \ref{tab:addlabel} for both datasets.

\end{document}